%% file: main.tex
\documentclass{sig-alternate-10pt}

\usepackage{times}
\usepackage{graphicx,cite}

\usepackage{amssymb,amsmath,amsbsy}

\usepackage{algorithm,algorithmic}

\usepackage{array}

\usepackage{balance}

\usepackage[hyphens]{url}

\def\squareforqed{\hbox{\rlap{$\sqcap$}$\sqcup$}}
\def\qed{\ifmmode\squareforqed\else{\unskip\nobreak\hfil
\penalty50\hskip1em\null\nobreak\hfil\squareforqed
\parfillskip=0pt\finalhyphendemerits=0\endgraf}\fi}

\newtheorem{lemma}{\bf{Lemma}}

\newcommand{\argmin}{\operatornamewithlimits{argmin}}

\begin{document}

\title{Reducing Electricity Demand Charge for Data Centers\\ with Partial Execution}

\numberofauthors{2} 
%
\author{
%
%
\alignauthor
Hong Xu\\
       \affaddr{Department of Computer Science}\\
       \affaddr{City University of Hong Kong}\\ 
       \affaddr{Kowloon, Hong Kong}\\
       \email{henry.xu@cityu.edu.hk}
\alignauthor
Baochun Li\\
       \affaddr{Department of Electrical and Computer Engineering}\\
       \affaddr{University of Toronto}\\
       \affaddr{Toronto, ON, Canada}\\
       \email{bli@eecg.toronto.edu}
}

\maketitle

\begin{abstract}

Data centers consume a large amount of energy and incur substantial electricity cost. In this paper, we study the familiar problem of reducing data center energy cost with two new perspectives. First, we find, through an empirical study of contracts from electric utilities powering Google data centers, that demand charge per kW for the maximum power used is a major component of the total cost. Second, many services such as Web search tolerate partial execution of the requests because the response quality is a concave function of processing time. Data from Microsoft Bing search engine confirms this observation.

We propose a simple idea of using partial execution to reduce the peak power demand and energy cost of data centers. We systematically study the problem of scheduling partial execution with stringent SLAs on response quality. For a single data center, we derive an optimal algorithm to solve the workload scheduling problem. In the case of multiple geo-distributed data centers, the demand of each data center is controlled by the request routing algorithm, which makes the problem much more involved. We decouple the two aspects, and develop a distributed optimization algorithm to solve the large-scale request routing problem. Trace-driven simulations show that partial execution reduces cost by 3\%--10.5\% for one data center, and by 15.5\% for geo-distributed data centers together with request routing.

\end{abstract}

\input{intro}

\input{motivations}

\input{model}

\input{onedc}

\input{evaluation}
\input{related}

\input{conclusion}

\section*{Acknowledgment}
We thank Yuxiong He from Microsoft Research Redmond for providing the response quality data from Bing, as well as insightful suggestions on the ideas of this paper. We also thank Minghua Chen from The Chinese University of Hong Kong, and Shaolei Ren from Florida International University for their encouragement and helpful discussions.
\bibliographystyle{abbrv}
\bibliography{IEEEabrv,main}

\end{document}

%% file: intro.tex
\section{Introduction}
\label{sec:intro}

Data centers are the powerhouse behind many Internet services today. A modern data center, deployed by companies such as Google, Microsoft, and Facebook, often hosts tens or even hundreds of thousands of servers to provide services for millions of users at the global scale \cite{google-servers, QWBG09}. Energy consumption of data centers is enormous: Google's data centers draw 260~MW of power in 2011 \cite{google-power}, and incur millions of dollars of electricity bills.

How to reduce data centers energy cost has thus received much attention over the recent years. Since servers and cooling systems constitute the majority of a data center's power budget \cite{ZWMB12}, reducing energy cost is commonly addressed on these two fronts. Workloads may be shifted across time and location to exploit the diversity of electricity prices \cite{QWBG09,RLXL10,LLWL11,LWAT11,GCWK12}. The cooling energy overhead can also be optimized with more efficient cooling systems and integrated thermal management \cite{BF07,FWB07,CGHW10,LCBW12,ZWMB12,icac13,sigmetrics13}. 

Despite extensive efforts, the fundamental question of how the electricity bill for data centers is actually calculated by utilities is not well understood. Almost all of the previous works simply assume that the cost is solely determined by the total energy consumption in kilowatt hours (kWh). We revisit this question by conducting empirical investigations. Data centers, like other large industrial power users, typically enter long-term contracts with local utilities instead of purchasing power off the market to avoid price volatility \cite{M13}. Thus, we collect real-world electricity contracts from utilities that power Google data centers, and study the pricing structures. 

Though details vary, we find that the electricity bill for data centers has two major components: energy charge, and demand charge. Energy charge is the commonly studied cost of total kWh used. Demand charge, on the other hand, calculates the cost of {\em peak} power used in kW during the billing period, and can be much more significant than energy charge. For example for a data center consuming 10~MW on peak and 6~MW on average, the monthly energy charge and demand charge amounts to around \$24,000 and \$165,500, respectively, according to Georgia Power's PLH-8 contract \cite{ga-contract}. How to reduce the demand charge, however, has not been fully discussed in the literature.

\begin{figure}[tph]
\centering
\includegraphics[width=0.46\textwidth]{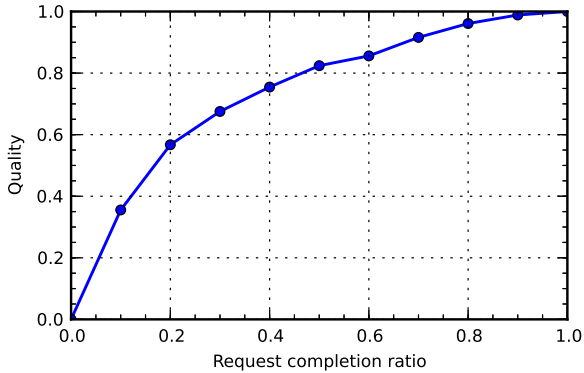}
\vspace{-4mm}
\caption{Search quality profile in Microsoft Bing search engine. Data is from 200K queries of a production trace \cite{HELY12}.}
\label{fig:quality}
\end{figure}

Motivated by this observation, in this paper, we advocate to reduce the peak power and demand charge of data centers, by using a simple idea of {\em partial execution}. Partial execution has been exploited to improve request completion time for interactive services \cite{HELY12}. Many interactive services execute tasks in a distributed and iterative fashion. Results of a user request will improve in quality given additional processing time and energy. The marginal improvement of response quality however is diminishing. A typical application that exhibits these properties is web search. Fig.~\ref{fig:quality} plots the empirical search quality profile from 200K queries in a production trace of Microsoft Bing \cite{HELY12}. The quality profile is clearly concave as a result of the diminishing marginal return in quality. Therefore, a request does not necessarily need to be fully executed: at peak hours, partial execution can be used to trade response quality for demand charge savings, in addition to improving the processing time.

The technical challenge of using partial execution is to develop efficient workload scheduling algorithms to decide when and how partial execution should be used, so that the demand charge and energy cost are minimized and the Service Level Agreement (SLA) on response quality is satisfied. Towards this end, we make the following contributions. 

First we propose a general optimization model to realistically capture both demand charge and energy charge according to our empirical study, and the typical percentile-based SLA constraints. We find that the SLA constraints imply that the optimal solution at each time slot is {\em binary}, i.e. we only need to decide whether to use a high power mode with high quality, or low power mode with low quality. This greatly simplifies the problem formulation. 

Our second contribution is a systematic study of the workload scheduling problem with partial execution. We consider the case of a single data center, where the problem is an integer program, and derive a simple optimal algorithm. We also study the case of multiple geo-distributed data centers, where each data center's request demand as well as power use can be adjusted by the request routing algorithm. This new dimension adds considerable complexity to solving the joint optimization in a practical manner (e.g., every 15 minute). We decouple the problem, by first optimizing request routing without partial execution to reduce the demand fluctuation seen by each data center, and then optimally solving workload scheduling with partial execution. The request routing problem itself is difficult due to its large scale and tight constraint coupling. We rely on the alternating direction method of multipliers (ADMM) that offers fast convergence \cite{BPCP10,BT97}. ADMM decomposes the problem into per-user and per-data center sub-problems that are easy to solve, leading to an efficient distributed algorithm.

Finally, we perform trace-driven simulations to evalute the cost reductions of partial execution with our algorithms in Sec.~\ref{sec:evaluation}. Results demonstrate that our algorithms outperform existing schemes that only focus on energy charge. A saving of 3\%--10.5\% can be realized for one data center depending on the relative importance of demand charge, and a saving of 15.5\% can be achieved for geo-distributed data centers.

%% file: motivations.tex
\section{Motivations }
\label{sec:motivations}

Let us start by motivating our key idea in this paper: using partial execution to reduce the energy cost, especially the demand charge of data centers. To make our case concrete, we first present an empirical analysis of the electricity billing method to demonstrate the importance of demand charge. We then introduce some background on the feasibility of partial execution for typical data center workloads, such as Web search.

\subsection{Electricity Billing: An Empirical Analysis of Contracts}
\label{sec:billing}

It is generally assumed that a simple volume-based charging scheme calculates the total energy cost for all kWh a data center consumes. Prices of the day-ahead or hour-ahead future markets operated by ISOs (Independent System Operator) or RTOs (Regional Transmission Operator) are often used as the prices data centers pay per kWh. However, ISOs and RTOs operate their markets mainly for electricity suppliers to balance the supply and demand of the grid in real time \cite{isorto}. Data centers, as an electricity consumer, do not participate in and purchase power off the ISO or RTO market. They generally enter long-term contracts with their local utilities to obtain fixed electricity prices and avoid volatility \cite{M13}. 

To see how a data center's electricity bill is calculated in practice, we perform an empirical analysis of real-world electricity long-term contracts, which to our knowledge has not been done before. We briefly explain our methodology here. Our study is based on the locations of all six Google data centers in the U.S. \cite{GoogleDCs}. We first determine the local utilities that power each of these data centers according to anecdotal evidence, as shown in Table~\ref{table:utilities}. In many cases there is only one electric utility operating in the region of a Google data center, which makes us believe that our inference is accurate.\footnote{We provide references to anecdotal evidence for determining the local utilities that power each Google data center in Table~\ref{table:utilities}. For those without references, they are the only utility in the region.} We then collect the long-term contracts these utilities provide for large industrial users---such as data centers---that has an annual contract demand of more than 10~MW. For simplicity we choose contracts with fixed rates instead of time-of-use pricing. All utilities in our study publish contracts and rate schedules on their websites, and all the contracts we study can be downloaded from \cite{contracts}. We believe that these contracts faithfully represent the billing method used in the actual contracts data centers enter. 
\begin{table*}
    \centering
    \small
    \caption{Electric utilities for Google data centers. Monthly cost breakdown based on a 10~MW peak demand and a 6~MW average demand. Data collected in June, 2013.}
    \begin{tabular}{| c | c | c | c | c |}
    \hline 
	    Location & Utility & Contract Type & Demand Charge & Energy Charge \\ \hline
        The Dalles, OR & Northern Wasco County PUD \cite{or} & Primary Service \cite{pud} & \$38,400 & \$147,312 \\
        Council Bluffs, IA & MidAmerican Energy & Large General Service, South System \cite{mid} & \$62,600 & \$114,236 \\
        Mayes County, OK & The Grand River Dam Authority \cite{ok} & Wholesale Power Service \cite{grda} & \$103,900 & \$93,312 \\
        Lenoir, NC & Duke Energy \cite{nc} & Large General Service LGS-24 \cite{duke} & \$111,000 & \$240,580 \\ 
        Berkeley County, SC & South Carolina Electric \& Gas Company & Rate 23 -- Industrial Power Service \cite{sceg} & \$147,600  & \$217,598 \\
        Douglas County, GA & Georgia Power & Power and Light -- High Load Factor PLH-8 \cite{gp} & \$165,500 & \$24,002 \\
    \hline  
    \end{tabular}
\label{table:utilities}
\end{table*}

Our empirical study reveals that the monthly electricity bill is, among other things, determined mainly by two methods: a volume-based charging method that charges the total kWh of energy used, and a peak-based charging scheme that charges the maximum demand measured in kW in the billing period. More specifically, utilities install demand meters at customers' facilities to record the average demand in kW for every 15 minutes in general. The customer is billed for the highest 15-minute average demand during the billing cycle. In the utilities' taxonomy, volume-based charging results in {\em energy charge}, and peak-based charging results in {\em demand charge}. Table~\ref{table:contract} shows the typical structure of a contract that we collected. 

\begin{table}
    \centering
    \small
    \caption{The Industrial Power Service contract, SCEG \cite{sceg}.}
    \begin{tabular}{| l | c | }
    \hline 
	    Item & Price (USD)  \\ \hline
	    Basic Facilities Charge & \$1925.00 \\
        Demand Charge & \$14.76/kW \\
	    Energy Charge & \$0.05037/kWh \\
	    Miscellaneous & Tax, minimum charge, etc. \\
    \hline  
    \end{tabular}
\label{table:contract}
\end{table}

Demand charge and energy charge constitute the majority of total energy cost. Intuitively, demand charge helps utilities recover the cost of providing capacity to meet the peak demand, which is more expensive than meeting the average demand especially for large industrial users. Thus, demand charge is in general on par with energy charge, and often significantly higher. We estimate the monthly cost breakdown of all utilities in Table~\ref{table:utilities}, for a typical data center that consumes 10~MW on peak and 6~MW on average. Observe that in the case of Georgia, demand charge is almost 8x energy charge. The importance of demand charge is more salient when the peak-to-average ratio of the demand increases.

Therefore, one needs to take into account demand charge in order to reduce energy cost, which unfortunately has not yet been fully explored. Previous works focus only on reducing the energy charge, that is the total energy consumption. They do not necessarily reduce the peak energy consumption and demand charge.

\subsection{Partial Execution: A Feasibility Check}
\label{sec:partial_bk}

We propose to exploit partial execution of requests to reduce both the peak and total energy consumption. Partial execution is orthogonal to, and can work with existing energy management approaches that focus on energy charge. We now provide a feasibility check for partial execution in the context of Web search, which is one of the most important data center workloads.

An Internet search engine consists of crawling, indexing, and query serving systems. We focus on the query serving system, which is a distributed system with many aggregators and index servers. When a query arrives and hits the cache, the results are immediately returned. Otherwise, it is assigned to an aggregator. The aggregator sends the request to index servers, each of which holds a partition of the entire index for billions of documents. An index server then searches its index for documents matching the keywords in the query. It ranks the matching documents sequentially using a PageRank-like algorithm. This is the most time- and energy-consuming part and it uses over 90\% of hardware resources \cite{HELY12}, because the ranking algorithm needs to extract and compare many features of the documents.

Web search is best-effort: The query response quality improves as more time and resources are used to run the ranking algorithm with more matching documents. Partial execution can be implemented in a rather straightforward way for a search engine, by setting a threshold for the ranking algorithm's running time. If the elapsed processing time reaches the threshold, the algorithm is terminated, and index servers return the top ranked results they compute. The quality profile as in Fig.~\ref{fig:quality} is concave, meaning that a small degree of partial execution will not severely impact quality. These observations thus confirm that partial execution is feasible in practice for data centers. 

In fact, besides Web search, many other systems also tolerate inexact or tainted results. For example it is acceptable to skip spelling correction when composing a complex web page with many sub-components \cite{D12}. Partial execution has already been adopted to rein in the tail request completion times in Google and Microsoft's Internet services \cite{D12, HELY12}. 


%% file: model.tex
\section{System Models}
\label{sec:model}

Before developing algorithms that control when and how partial execution should be used to save cost, we first state our models and assumptions in this section.

We consider a discrete time model, where in each slot $t$ the average power draw is measured at the data center. There is an interval of interest $t\in\{1, \ldots, T\}$. The length of a time slot equals 15 minutes, and the planning horizon $T$ is one day ($T=240$) in which the demand series can be accurately predicted. This is a valid assumption in practice. Time series analysis and other learning algorithms have been shown to fairly accurately predict the aggregate demand from a large number of users, which exhibits regular patterns \cite{GGW10,NXLZ12}. We cannot consider a longer planning horizon, say one month, for which prediction becomes unreliable. However we show through simulations in Sec.~\ref{sec:evaluation} that our algorithms perform close to the ideal case when we have limited future information. The partial execution decision is adjusted every 15 minutes for all requests.

\subsection{Server Power and Energy Cost}
\label{sec:power_model}

We adopt the empirical model from \cite{FWB07} that calculates the individual server power consumption as an affine function of CPU utilization at $t$, $E_{\text{I}}+\left( E_{\text{P}} - E_{\text{I}} \right)u(t)$. $E_{\text{I}}$ is the server power when idle, $E_{\text{P}}$ is the server power when fully utilized, and $u(t)$ is the average CPU load at $t$. This model is especially accurate for calculating the aggregate power of a large number of servers \cite{FWB07}. 
$u(t)$ is determined by the 15-minute request demand $D(t)$, and the request completion ratio $\alpha(t)\in[0,1]$ which we control. Assuming the data center deploys $N$ index servers to process search queries, the cache miss rate is 10\%, and each request takes 50~ms to complete with 200 servers running at 100\% CPU utilization, we have:
\begin{equation*}
	u(t) = D(t)\cdot0.1\cdot 200\cdot 0.05\cdot \alpha(t) / N\cdot 15\cdot 60 = \alpha(t)D(t)/900N 
\end{equation*}
We assume that servers are adequately provisioned and demand can always be handled so that $u(t)\le 1$ holds, i.e.,
\begin{equation}
	N\ge D(t)/900, \forall t. \label{eqn:N}
\end{equation}

Since servers are always on once commissioned \cite{GHMP09,M13}, server idle power is an immaterial constant that we do not consider subsequently. The total server usage power in kW at $t$ is then a linear function of both $\alpha(t)$ and $D(t)$:
\begin{equation}\label{eqn:server_power}
	E\big( \alpha(t), D(t)\big) = (E_{P} - E_{I})\frac{D(t)\alpha(t)}{900}. 
\end{equation}

As discussed in Sec.~\ref{sec:billing}, the electricity bill has both demand charge and energy charge. Denote the demand price as $P^D$ per kW, and the energy price as $P^E$ per 0.25 kWh (recall a time slot is 0.25 hour). The total energy cost is then:
\begin{equation}\label{eqn:energy_cost}
	\max_{t\in[1,T]} E\big( \alpha(t), D(t) \big)P^D  + \sum_{t=1}^T  E\big( \alpha(t), D(t) \big)P^E  
\end{equation}

Since we use partial execution for Web search, we are only concerned with the energy consumption of index servers. Other components of the infrastructure, such as the cooling system, also consume a lot of power \cite{ZWMB12}. They can be accounted for by a multiplying PUE factor to the server power, which captures the energy overhead as a function of the ambient temperature, humidity, etc. \cite{GCWK12,icac13}, without fundamentally changing the nature of our problem. Thus we do not model them in this paper.


\subsection{SLA on Response Quality}
\label{sec:sla}

For a search engine, response quality is arguably one of the most important performance metrics. Response quality here compares the tainted results of partial execution against those from full execution. Thus many commercial services specify strict Service Level Agreements (SLAs), using both high-percentile and worst-case response quality. 
High-percentile guarantees ensure consistent high-quality results, at the extremes of the service distribution. For example, a web search may have an SLA that targets a 0.99 quality for at least 95\% of requests, referred to as the 95$^{\text{th}}$-percentile quality \cite{HELY12}. Worst-case guarantees, e.g. at least a 0.8 quality needs to be met for all requests, ensure that performance is at an acceptable level as the bottom line. 

We model SLAs using the empirical response quality profile of Bing as shown in Fig.~\ref{fig:quality}. Specifically, response quality is a function of the request completion ratio $Q(\alpha) \in[0,1]$. $Q(\alpha)$ can be obtained by applying regression techniques to the empirical data points in Fig.~\ref{fig:quality}:
\begin{equation}\label{eqn:quality}
	Q ( \alpha ) = -0.82129975 \alpha^2 + 1.67356677 \alpha +0.14773298.
\end{equation}
Clearly $Q(\alpha)$ is concave in $[0,1]$. To save cost, the operator will use just enough resources to satisfy the SLA. In other words, the operator makes sure that the quality of 95\% of the requests is exactly 0.99, and the quality of the rest 5\% requests is exactly 0.8 according to our examples above. Thus, in the problem of minimizing energy cost with demand charge, the partial execution decision is {\em binary}, even though the entire range from 0 to 1 is possible to implement. At each time slot, the operator needs to make a decision of whether to operate in the high power mode where the response quality is 0.99, or in the low power mode where quality is 0.8.

This observation greatly simplifies our model. We let $X(t)$ be a binary indicator of the partial execution decision at each time slot $t$. $X(t)=1$ if $Q\big( \alpha(t) \big) = 0.99$, i.e. $\alpha(t)=Q^{-1}(0.99)$, and $X(t)=0$ if $Q\big(\alpha(t) \big) = 0.8$, i.e. $\alpha(t) = Q^{-1}(0.8)$. It is then only necessary to make sure that the 95$^{\text{th}}$-percentile quality guarantee is satisfied, which amounts to the following:
\begin{equation}\label{cons:sla}
	\sum_{t=1}^{T} X(t)D(t) \ge 0.95 \sum_{t=1}^{T} D(t).
\end{equation}

At this point, some may wonder to what extent could partial execution reduce cost. After all, only 5\% of the requests can be served using the low power mode, and they still need to have a 0.8 quality. Notice that since $Q(\alpha)$ is concave, a 0.8 quality can cut the processing time by half from \eqref{eqn:quality}, which implies a good amount of power reduction. Also demand charge can be reduced substantially by only using partial execution at a few time slots. Our claims will be verified in Sec.~\ref{sec:evaluation}.

%% file: onedc.tex
\section{Algorithms}
\label{sec:algorithms}

We now systematically study the data center workload scheduling problem with partial execution. We formally introduce the problem formulations and solution algorithms in the cases of both a single data center and multiple geo-distributed data centers.

\subsection{The Case of One Data Center}
\label{sec:onedc}

As a starting point, we consider one data center. At $t=1$, the demand information $\{ D(t) \}$ is known for the interval $T$. Given $\{D(t)\}$, we can formulate the problem as:
\begin{align}\label{opt:onedc}
	\min\; &  \max_{t\in[1,T]} E\big( \alpha(t), D(t) \big)P^D   + \sum_{t=1}^T  E\big( \alpha(t), D(t) \big)P^E  \nonumber \\
	\text{s.t. } & \sum_{t=1}^{T} X(t)D(t) \ge 0.95 \sum_{t=1}^{T} D(t), \nonumber \\
	& \alpha(t) = \left\{ \begin{array}{cl}
		Q^{-1}(0.99), & \text{if } X(t)=1, \\
		Q^{-1}(0.8), & \text{if } X(t)=0. 
		\end{array}\right. \nonumber \\
	\text{variables: } & X(t), \forall t.
\end{align}
The workload schedule $\{ X(t) \}$ are our optimizing variables. They entail whether the high or the low power mode should be used at each time slot. In words, our problem is to determine the optimal workload schedule that minimizes the total cost for the period while conforming to the SLA.

The optimization \eqref{opt:onedc} is an integer program, which is hard to solve in general. A moment's reflection tells us that it is not the case for our problem. Since we know the demand series, and setting $X(t)=0$ reduces both terms of the objective function, we can derive the optimal solution with a trial-and-error approach summarized in Algorithm~\ref{alg:onedc}. We initialize all $X(t)$ to $1$. In the decreasing order of demand, the algorithm goes through all time slots. For each $t$, it sets each $X(t)=0$ if this does not violate the SLA, and reverts $X(t)$ to $1$ if otherwise.
\begin{algorithm}
	\caption{\textsl{Optimal Solution for \eqref{opt:onedc}}} 
\algsetup{linenosize=\small, linenodelimiter=.}
\begin{algorithmic}[1]
	\STATE{Initialize $X(t)$ to $1$ for all $t$. }
	\WHILE{$\{D(t)\}$ is not empty}
	\STATE{Pick the highest $D(t)$, and set $X(t)=0$.\label{step:trial} }
	\IF{\eqref{cons:sla} holds with $X(t)=0$}
		\STATE{Output $X(t)=0$.}
	\ELSE
		\STATE{Output $X(t)=1$.}
	\ENDIF
	\STATE{Set $D(t) = 0$. }
	\ENDWHILE
\end{algorithmic}
\label{alg:onedc}
\end{algorithm}

The optimality of the workload schedule is intuitive. The solution is feasible to problem \eqref{opt:onedc} for at each step we ensure the SLA is satisfied. It is also optimal since we always set the most demanding time slots in low power mode whenever possible, thereby providing the largest cost reduction in both demand and energy charge. 

\subsection{The Case of Geo-distributed Data Centers}
\label{sec:geo}

We have assumed a single data center, in which case the problem can be solved relatively easily. In practice we may have multiple data centers geographically distributed over the wide area to improve the service latency and reliability. In this case, the provider deploys some mapping nodes, such as authoritative DNS servers or HTTP ingress proxies \cite{WJFR10,NSS10}, to route user requests to an appropriate data center based on certain criteria. Thus an individual data center's demand is determined by the request routing algorithm. Request routing has been studied in many recent works to exploit price diversity and save energy charge \cite{QWBG09,LLWL11,GCWK12,XL13,icac13}. Yet it has not been studied with demand charge, where the routing decision needs to smooth out the demand series for each data center. 

We consider a provider with $J$ data centers, each running $N_j$ index servers. In the subsequent analysis the same subscript $j$ is appended to all the notations introduced in Sec.~\ref{sec:model} to denote the location specific quantities when necessary. We allow a mapping node to arbitrarily split a user's request traffic among all data centers. DNS servers and HTTP proxies can achieve such flexibility in commercial products \cite{WJFR10,KMSJ09}. Let ${I}$ denote the number of users. In this work a user $i$ is simply a unique IP prefix similar to \cite{NSS10}. Now at each time slot, the operator computes the request routing decisions together with workload schedules to better cope with dynamic request demand and reduce cost.  
The joint problem can be formulated as follows:
\begin{align}
	\min\; &  \sum_{j=1}^{J}\max_{t\in[1,T]} E_j\left( \alpha_j(t), \sum_{i=1}^I d_{ij}(t) \right)P^D_j \nonumber \\ 
	& + \sum_{j=1}^{J}\sum_{t=1}^T  E_j\left( \alpha_j(t), \sum_{i=1}^I d_{ij}(t) \right)P^E_j \nonumber \\
	\text{s.t. } & \sum_{t=1}^{T} X_j(t) \sum_{i=1}^I d_{ij}(t) \ge 0.95 \sum_{t=1}^{T} \sum_{i=1}^I d_{ij}(t),\forall j, \nonumber \\
	& \sum_{j=1}^J d_{ij}(t) = D_i(t), \forall i, t, \label{cons:workload} \\
	& \sum_{j=1}^J d_{ij}(t) L_{ij} / D_i(t) \le L, \forall i, t, \label{cons:latency} \\
	& \sum_{i=1}^I d_{ij}(t) \le 900 N_j, \forall j, t, \label{cons:capacity} \\
	& \alpha_j(t) = \left\{ \begin{array}{cl}
		Q^{-1}(0.99), & \text{if } X_j(t)=1, \\
		Q^{-1}(0.8), & \text{if } X_j(t)=0.
		\end{array}\right. \forall j. \nonumber \\
	\text{variables: } & X_j(t), d_{ij}(t), \forall i,j, t.\label{opt:geodc}  
\end{align}

We use $d_{ij}(t)$ to denote the amount of requests routed to data center $j$ from user $i$ at $t$. $\{d_{ij}(t)\}$ and $\{X_j(t)\}$ are our decision variables.
Compared to \eqref{opt:onedc}, the objective function is now the sum of costs from all data centers, which can be optimized by both the workload schedule and the request routing decision. There are three additional constraints: \eqref{cons:workload} is a user workload conservation constraint that requires a user's demand to be fully satisfied at all times; \eqref{cons:latency} is a user latency constraint that states a user's average latency should not be worse than $L$; and \eqref{cons:capacity} is the simple data center capacity constraint as discussed in Sec.~\ref{sec:power_model}. $L_{ij}$ denotes the end-to-end network latency between user $i$ and $j$, which can be obtained through active measurements. 

The optimization is a mixed-integer program (MIP) with a convex objective function. Adding to the complexity of the problem is its large scale. The number of users $I$, i.e. unique IP prefixes, can be $O(10^5)$ for a production cloud. The number of data centers $J$ is $O(10)$, and the number of time slots $T=240$. Thus \eqref{opt:geodc} has $O(10^8)$ variables, and $O(10^6)$ constraints. This prohibits a direct approach of using an optimization package to solve the problem, as it takes more than 15 minutes for a modern solver to solve MIPs with millions of variables and constraints \cite{M11}.

Since directly solving the joint optimization is infeasible, we decouple the request routing and workload scheduling problem to reduce the complexity. Specifically, we first solve the request routing problem and obtain the solution $d^*_{ij}(t)$ without partial execution, by setting all $X_{j}(t)$ to $1$. We then solve the workload scheduling problem using Algorithm~\ref{alg:onedc} for each data center $j$ given the demand $\sum_i d^*_{ij}(t)$. Though sub-optimal, this approach still allows request routing to effectively smooth out the demand peaks seen by data centers in the worst case.

Thus from now on we focus on solving the decoupled request routing problem with all $X_j(t)=1$. Since the server power function $E_j(\cdot)$ as in \eqref{eqn:server_power} is linear in both $\alpha_j(t) $ and $d_{ij}(t) $, the decoupled request routing problem can be written as:
\begin{eqnarray}
	\min\; &  \displaystyle\sum_{j=1}^{J}\max_{t\in[1,T]} E_j\left( \sum_{i=1}^I d_{ij}(t) \right)P^D_j \nonumber \\ 
	 & + \displaystyle\sum_{j=1}^{J}\sum_{t=1}^T   \sum_{i=1}^I E_j\big(d_{ij}(t) \big)P^E_j \nonumber \\
	\text{s.t. } & \eqref{cons:workload}, \eqref{cons:latency}, \eqref{cons:capacity}. \label{opt:routing}
\end{eqnarray}
Note $\alpha_j(t)$ can now be omitted in $E_j(\cdot)$ without loss of generality. This is a large-scale convex optimization which still has $O(10^8)$ variables and $O(10^6)$ constraints. More importantly, the constraints \eqref{cons:workload}, \eqref{cons:latency} and \eqref{cons:capacity} couple all variables together, which makes it difficult to solve. The coupling here in this case is especially difficult, because it happens on two orthogonal dimensions simultaneously: The per-user constraints \eqref{cons:workload} and \eqref{cons:latency} couple $d_{ij}(t)$ across data centers, and the per data center capacity constraint \eqref{cons:capacity} couples $d_{ij}(t) $ across users.

In these cases we rely on a distributed algorithm that enables parallel computations in data centers. A common approach is to relax the constraints and employ dual decomposition to decompose the problem into many independent sub-problems \cite{CLCD07}. Subgradient methods can then be used to update dual variables towards the optimum of the dual problem \cite{BM}. Yet, dual decomposition does not apply here, because it requires the objective function to be strictly convex, for otherwise the Lagrangian is unbounded below. Our objective function, including a $\max$ and a linear function, is not strictly convex.

Summarizing the discussions, we need to design a practical distributed algorithm that does not require strict convexity of the objective function, and preferably converges fast for large-scale problems. Next, we present such an algorithm based on the alternating direction method of multipliers (ADMM) \cite{BPCP10} .

\subsection{A Distributed Request Routing Algorithm}
\label{sec:admm}

We first provide a brief primer on ADMM. Developed in the 1970s \cite{BT97}, ADMM has recently received renewed interest in solving large-scale distributed convex optimization in statistics, machine learning, and related areas \cite{BPCP10}. The algorithm solves problems in the form
\begin{eqnarray}\label{opt:admm}
	\min & f(x) + g(z)\\
	\text{s.t. } & Ax + Bz = c, \nonumber\\
				& x \in C_1, z \in C_2, \nonumber
\end{eqnarray}
with variables $x\in\mathbf{R}^n$ and $z\in\mathbf{R}^m$, where $A\in\mathbf{R}^{p\times n}$, $B\in\mathbf{R}^{p\times m}$, and $c\in\mathbf{R}^{p}$. $f$ and $g$ are convex functions, and $C_1, C_2$ are non-empty polyhedral sets. Thus, the objective function is {\em separable} over two sets of variables, which are coupled through an equality constraint.

We can form the {augmented Lagrangian} \cite{H69} by introducing an extra $\mathcal{L}$-$2$ norm term $\|Ax+Bz-c\|^2_2$ to the objective:
\begin{multline}\label{eqn:aug_lag}
	L_\rho(x,z,\lambda) = f(x) + g(z)+\lambda^T(Ax+Bz-c)\\+(\rho/2)\|Ax+Bz-c\|^2_2.
\end{multline}
$\rho>0$ is the penalty parameter ($L_0$ is the standard Lagrangian for the problem). The augmented Lagrangian can be viewed as the unaugmented Lagrangian associated with the problem
\begin{eqnarray*}
	\min & f(x) + g(z) + (\rho/2)\|Ax+Bz-c\|^2_2\\
	\text{s.t. } & Ax + Bz = c, \nonumber\\
				& x \in C_1, z \in C_2. \nonumber
\end{eqnarray*}
Clearly this problem is equivalent to the original problem \eqref{opt:admm}, since for any feasible $x$ and $z$ the penalty term added to the objective is zero. The benefit of the quadratic penalty term is that it makes the objective function strictly convex for all $f$ and $g$. The penalty term is also called a regularization term and it helps substantially improve the convergence of the algorithm.

ADMM solves the dual problem with the iterations:
\begin{align}
	x^{t+1} &:= \argmin_{x\in C_1} L_\rho(x,z^t,\lambda^t), \label{eqn:x-min}\\
	z^{t+1} &:= \argmin_{z\in C_2} L_\rho(x^{t+1},z,\lambda^t), \label{eqn:z-min}\\
 \lambda^{t+1} &:= \lambda^t + \rho(Ax^{t+1}+Bz^{t+1}-c).\label{eqn:dual-update}
\end{align}
It consists of an $x$-minimization step \eqref{eqn:x-min}, a $z$-minimization step \eqref{eqn:z-min}, and a dual variable update \eqref{eqn:dual-update}. Note the step size is simply the penalty parameter $\rho$. Thus, $x$ and $z$ are updated in an alternating or sequential fashion, which accounts for the term {\em alternating direction}. Separating the minimization over $x$ and $z$ is precisely what allows for decomposition when $f$ or $g$ are separable, which will be useful in our algorithm design.

The optimality and convergence of ADMM can be guaranteed under very mild technical assumptions.
For more details about convergence see \cite{BT97}. In practice, it is often the case that ADMM converges to modest accuracy within a few tens of iterations \cite{BPCP10}, which makes it attractive for practical use.

Our request routing problem \eqref{opt:routing} cannot be readily solved using ADMM. The constraints \eqref{cons:workload} and \eqref{cons:capacity} couple all variables together as mentioned before, whereas in ADMM problems the constraints are separable for each set of variables. However, in spirit, our problem is close to the general ADMM form \eqref{opt:admm}.

To address this, we introduce a new set of {auxiliary variables} $ b_{ij}(t) = d_{ij}(t) $, and re-formulate the problem:
\begin{eqnarray}
	\min_{d,b}\; &  \displaystyle\sum_{j=1}^{J} \max_{t\in[1,T]} E_j\left( \sum_{i=1}^I d_{ij}(t) \right)P^D_j \nonumber \\ 
	 & + \displaystyle\sum_{j=1}^{J} \sum_{t=1}^T  \sum_{i=1}^I E_j\big( b_{ij}(t) \big)P^E_j \nonumber \\
	\text{s.t. } & b_{ij}(t) = d_{ij}(t), \forall i, j, t, \nonumber \\
	& \displaystyle\sum_{i=1}^I d_{ij}(t) \le 900 N_j, \forall j, t, \nonumber\\
	& \displaystyle\sum^J_{j=1} b_{ij}(t) = D_i(t), \displaystyle\sum_{j=1}^J b_{ij}(t) L_{ij} / D_i(t) \le L, \forall i, t.\label{opt:new_routing}
\end{eqnarray}
This technique is reminiscent to \cite{XL13}. This problem \eqref{opt:new_routing} is clearly equivalent to the original problem \eqref{opt:routing}. Observe that the new formulation is in the ADMM form \eqref{opt:admm}. The objective function is now separable over two sets of variables $d_{ij}(t) $ and $b_{ij}(t) $. $d_{ij}(t) $ controls the demand charge, while $b_{ij}(t) $ determines the energy charge. $d_{ij}(t) $ and $b_{ij}(t) $ are connected through an equality constraint. Overall, they control the provider's total energy cost of running the index servers. 

The use of auxiliary variables also enables the separation of per-user and per-data center constraints, and is the key step towards reducing the complexity as we demonstrate now. The augmented Lagrangian of \eqref{opt:new_routing} is
\begin{multline}\label{eqn:our_lag}
	\hspace{-4mm} L_\rho(d,b,\lambda) = \sum_j \max_{t\in[1,T]} E_j\left( \sum_{i=1}^I d_{ij}(t) \right)P^D_j + \sum_{i,j,t} E_j\big( b_{ij}(t) \big)P^E_j \\  
		+\sum_{i,j,t} \left(\lambda_{ij}(t)\big( d_{ij}(t) - b_{ij}(t) \big)+\frac{\rho}{2}\big(d_{ij}(t) - b_{ij}(t)\big)^2\right),
\end{multline}
where $d, b, \lambda$ are shorthands for $\{d_{ij}(t) \}, \{b_{ij}(t) \}, \{\lambda_{ij}(t) \}$.

The dual problem is solved by updating $d$ and $b$ sequentially. At the $k$-th iteration, the $d$-minimization step tries to minimize $L_{\rho}(d, b^{k-1}, \lambda^{k-1}) $ over $d$ with the capacity constraints \eqref{cons:capacity} according to \eqref{eqn:x-min}. By inspecting \eqref{eqn:our_lag}, we can readily see that this is {\em decomposable} over data centers since all terms related to $d$ are separable over $j$. Effectively, each data center needs to independently solve the following sub-problem:
\begin{align}
	\hspace{-1.5mm}\min_{d} & \max_{t\in[1,T]} E_j\left( \sum_{i=1}^I d_{ij}(t) \right)P^D_j \nonumber \\
	& + \sum_{i,t} d_{ij}(t) \left( \lambda^{k-1}_{ij}(t) + \frac{\rho}{2}\big(d_{ij}(t) - b^{k-1}_{ij}(t)\big) \right) \nonumber \\
	  \text{s.t. } & \sum^I_{i=1}d_{ij}(t) \le 900 N_j, \forall t. \label{opt:per-dc} 
\end{align}
The physical meaning of the per-data center problem is simple. Each data center computes the optimal request routing solution $d$ that minimizes the sum of its demand charge and the penalty of violating the constraint $d = b^{k-1}$. In other words, the data center also takes into account the users' perspective of the problem represented by $b^{k-1}$, eventually making sure that both parties converge to the same global optimal solution.

The per-data center sub-problem is a much simpler convex problem with $O(10^7)$ variables and only $T=O(10^2)$ constraints. Since the constraints are not coupled across multiple dimensions as in \eqref{opt:routing}, it can now be efficiently solved using a standard optimization solver. 

We have solved the $d$-minimization step distributively across all data centers by decomposing into $J$ per-data center subproblem in the form \eqref{opt:per-dc}. After obtaining the solution $d^{k}$, the $b$-minimization step can also be similarly attacked.

According to \eqref{eqn:z-min}, the $b$-minimization step tries to minimize $L_{\rho}(d^k, b, \lambda^{k-1}) $ over $b$ with the workload conservation constraints $\sum_j b_{ij}(t) = D_i(t),\forall i,t $. Readily it can be seen that this can also be decomposed across users, where each user independently solves the following per-user sub-problem:
\begin{align}
	\min_{b}\ & \sum_{j} \bigg( E_j\big( b_{ij}(t) \big)P^E_j + \frac{\rho}{2} b^2_{ij}(t) + \big(\rho d^k_{ij}(t) - \lambda^{k-1}_{ij}(t)\big)b_{ij}(t)\bigg) \nonumber \\
	 \text{s.t. } & \sum_j b_{ij}(t) = D_i(t), \sum_j b_{ij}(t)L_{ij}/D_i(t) \le L, \label{opt:per-user} 
\end{align}
which a simple quadratic program for $E_j()$ is linear. Again, the formulation embodies an intuitive interpretation. Here user $i$, at each $t$, optimizes its request routing strategy $\{b_{ij}(t)\}$ according to the prices $\{P^E_j\}$ to minimize the energy charge. Meanwhile, it also considers the data center's optimal solution that mainly concerns the demand charge, by staying close to $d^k_{ij}(t) $ and minimizing the quadratic penalty term as much as it can.

Having obtained the optimal $d^{k}$ and $b^{k}$, the final step is to perform the dual variable update:
\begin{equation}\label{eqn:lambda}
	\lambda^{k}_{ij} = \lambda^{k-1}_{ij} + \rho(d^{k}_{ij} - b^{k}_{ij}).
\end{equation}
The entire procedure is summarized in Algorithm~\ref{alg:ours}. Since the constraint set for $d$ is clearly bounded in our problem, according to \cite{BT97} the algorithm converges to the optimal solution.  

\begin{lemma}\label{lem:convergence}
	Our algorithm based on ADMM converges to the optimal solution $d^*$ and $b^*$ of \eqref{opt:new_routing} and equivalently \eqref{opt:routing}.
\end{lemma}

\begin{algorithm}
	\caption{\textsl{Optimal Distributed Solution for \eqref{opt:routing}}} 
\algsetup{linenosize=\small, linenodelimiter=.}
\begin{algorithmic}[1]
	\STATE{Initialize $d^0=0$, $b^0=0$, $\lambda^0=0$, $\rho=1$.}
	\STATE{At $k$-th iteration, solve $J$ per-data center sub-problems \eqref{opt:per-dc} in parallel. Obtain $d^k$.}\label{step:solving_alpha}
	\STATE{Given $d^{k}$, solve $I\cdot T$ per-user sub-problems \eqref{opt:per-user} in parallel. Obtain $b^k$.}\label{step:solving_beta}
	\STATE{Update dual variables $\lambda^k$ as in \eqref{eqn:lambda}.}
	\STATE{Return to step \ref{step:solving_alpha} until convergence.}
\end{algorithmic}
\label{alg:ours}
\end{algorithm}

Now to summarize, our algorithm follows a divide-and-conquer paradigm. Recall that $d$ controls the demand charge of processing the requests, while $b$ determines the energy charge. Our algorithm separately optimizes $d$ and $b$ for either aspect of the problem. Additionally, the penalty terms (i.e. the Augmented Lagrangian) force $d$ and $b$ to stay close to each other, eventually ensuring that they converge to the same request routing solution which is also optimal.

\subsection{Implementation Issues of Algorithm~\ref{alg:ours}}
\label{sec:implementation}

The distributed nature of Algorithm~\ref{alg:ours} allows for an efficient parallel implementation in a data center that has abundant server resources. Here we discuss several issues pertaining to such an implementation in reality. 

First, at each iteration, step~\ref{step:solving_alpha} can be implemented on $J$ servers, each solving one instance of the large-scale per-data center sub-problem \eqref{opt:per-dc}. Step~\ref{step:solving_beta} can be implemented even on a single server since it only involves solving quadratic programs \eqref{opt:per-user}. A multi-threaded implementation can further speed up the algorithm on multi-core hardware. Thus only $J$ servers are required to run the distributed algorithm.

Second, our algorithm can be terminated before convergence is reached. This is because ADMM is not sensitive to step size $\rho$, and usually finds a solution with modest accuracy within tens of iterations \cite{BPCP10}. A solution with modest accuracy is sufficient in situations of flash crowds of requests and failure recovery. The operator can apply an early-braking mechanism in these cases to terminate the algorithm after several tens of iterations without much performance loss.

Finally, the message passing overhead of our algorithm is also low. The request routing decisions $d$ need to be disseminated to the mapping nodes and data centers. All the other message passing, for exchanging $d$, $b$, and $\lambda$ amongst servers, happens in the internal network of the designated data center, which in many cases is specifically designed to handle the broadcast and shuffle transmission patterns of HPC applications such as MapReduce \cite{ALV08}. The amount of intermediate data our algorithm produces is much smaller than the bulky data of HPC applications \cite{VPSK09}. Thus the message passing overhead incurred to the data center network is low.

%% file: evaluation.tex
\section{Evaluation}
\label{sec:evaluation}

To realistically evaluate the cost reduction of partial execution with our algorithms, we conduct trace-driven simulations in this section.

\subsection{Setup}
\label{sec:eva_setup}

\begin{figure}[h]
	\centering
	\includegraphics[width=0.45\textwidth]{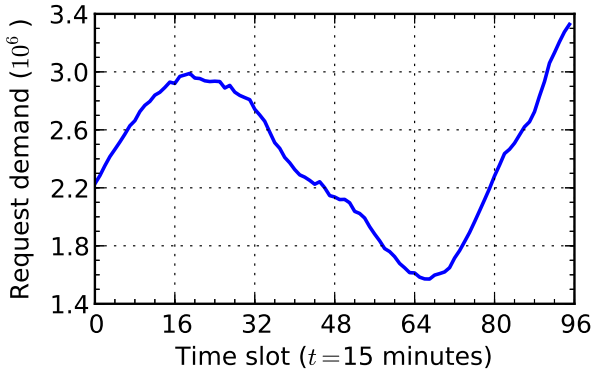}
	\vspace{-3.5mm}
	\caption{Total request traffic of the Wikipedia traces \cite{UPS09}.}
	\label{fig:demand}
\end{figure}

We use the Wikipedia request traces \cite{UPS09} to represent the Web search request traffic of a data center. The dataset we use contains, among other things, 10\% of all user requests issued to Wikipedia from a 30-day period of September 2007. 
The prediction of workload can be done accurately as demonstrated by previous work, and in the simulation we simply adopt the measured request traffic as the total demand. The scheduling period is 15 minutes, and the planning horizon $T$ is one day as mentioned in Sec.~\ref{sec:model}. Fig.~\ref{fig:demand} plots the request traffic of the traces for 24 hours of the measurement period. The scale of the traces closely matches Google's search traffic, which is roughly 1.2 trillion annual searches in 2012 \cite{gsearch}, or equivalently 2.7 million searches per 15 minutes per data center with its 13 data centers \cite{GoogleDCs}.

We consider six Google data centers in the U.S. We scale the Wikipedia traffic trace by a factor of six, and time shift it according to the time differences of these locations to synthesize the total demand of the six data centers. In the case of a single data center, the original trace is used. We rely on iPlane \cite{MIPD06}, a system that collects wide-area network statistics from Planetlab vantage points, to obtain the latency information. We set the number of clients $|\mathcal{I}|=10^5$, and choose $10^5$ IP prefixes from a RouteViews \cite{routeviews} dump. We then extract the corresponding round trip latency information from the iPlane logs, which contain traceroutes made to a large number of IP addresses from Planetlab nodes. We only use latency measurements from Planetlab nodes that are close to our data center locations. Since the Wikipedia traces do not contain any client information, to emulate the geographical distribution of requests, we split the total request traffic among the clients following a normal distribution.

Each data center has $N=N_j=5,000$ index servers\footnote{A data center has more than just index servers. Here we focus on index servers with partial execution.}, so it can process 4.5 million requests every 15 minutes according to \eqref{eqn:N}, while the peak demand of our trace is about 3.4 million requests. We use the contract prices of the local electric utilities that power these Google data centers as detailed in Sec.~\ref{sec:billing}. We assume a server's idle and peak power are $E_I=400$~W and $E_P=750$~W, respectively, which are typical for a data center server \cite{VSSS10}.

\begin{figure}[h]
	\centering
	\includegraphics[width=0.45\textwidth]{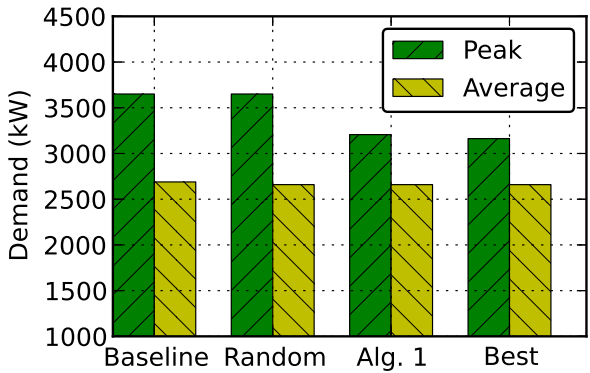}
	\vspace{-3.5mm}
	\caption{Monthly power consumption comparison for one data center.}
	\label{fig:d-comparison}
\vspace{-3mm}
\end{figure}

\subsection{The Case of One Data Center}
\label{sec:eva_onedc}

We start with one data center, and evaluate the benefit of partial execution with Algorithm~\ref{alg:onedc}. We solve the workload scheduling problem \eqref{opt:onedc} on a daily basis for the 30-day period, to obtain the monthly bill. We compare the performance of Algorithm~\ref{alg:onedc}, called {\sf Alg. 1}, with three benchmarks. The first one, called {\sf Baseline}, is a naive approach that does not use partial execution, and the data center is always operating in the high power mode. The second one, called {\sf Random}, uses partial execution randomly without our workload scheduling algorithm. This represents state-of-the-art that exploits partial execution for improving latency while satisfying SLA \cite{HELY12}, instead of using it to reduce the demand charge. The third one, called {\sf Best}, assumes that complete demand information for the entire 30-day period is known, and uses Algorithm~\ref{alg:onedc} to obtain the optimal schedule with minimum cost. This benchmark helps us understand the impact of limited future knowledge about the workload demand on reducing energy cost. 

\begin{figure}[hpt]
\centering
\includegraphics[width=0.45\textwidth]{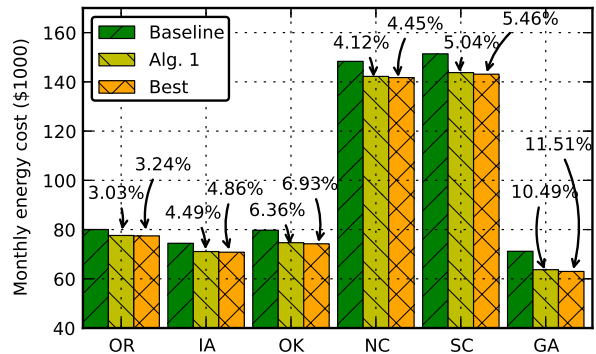}
\vspace{-4mm}
\caption{Monthly energy cost comparison for one data center.}
\vspace{-3mm}
\label{fig:cost-comparison}
\end{figure}

Fig.~\ref{fig:d-comparison} plots the monthly power consumption breakdowns, including both the peak and average demand, for the three benchmarks. Note that this calculation includes server idle power. All schemes reduce the average demand by 5\% compared to {\sf Baseline}, which is the maximum that the SLA allows. First notice that {\sf Random} only marginally reduces the peak power demand by 0.02\%, since it does not utilize partial execution strategically at times when demand is high. Our Algorithm~\ref{alg:onedc} utilizes limited (1-day) information that is practically available, and optimizes the partial execution schedule. Thus it is able to reduce the peak demand more substantially than {\sf Random} by 12.17\%. Also observe that when we have perfect future knowledge, {\sf Best} reduces the peak demand by 13.36\%, only slightly higher than {\sf Alg. 1}. This demonstrates that limited future knowledge provides close-to-optimal peak reduction with partial execution.

Fig.~\ref{fig:cost-comparison} shows the monthly energy cost comparison by using the contract prices of all six utilities in order to understand how much cost saving our idea can offer. Clearly we see that given the same demand series and partial execution schedules the total cost varies wildly depending on the prices. NC and SC are the most expensive locations while others are much cheaper. In all cases, {\sf Alg. 1} offers 3.04\% to 10.49\% total cost reductions compared to {\sf Baseline} without partial execution, and is again very close to {\sf Best}. The improvement becomes more salient for locations where demand charge is more significant than energy charge, such as OK and GA. In dollar terms, cost savings range from about \$2,400 to \$7,600 per month. Though the amount seems small for a data center, with the rapid increase of user demand and energy cost even a single digit of cost saving is crucial for operators. Moreover, an operator usually deploys multiple data centers, in which case the cost savings multiply and become more substantial even without optimizing request routing.

\subsection{The Case of Geo-distributed Data Centers}
\label{sec:eva-geodc}
 
\begin{figure*}[tbp]
	\begin{minipage}[t]{0.33\linewidth}
	\centering
	\includegraphics[width=1\linewidth]{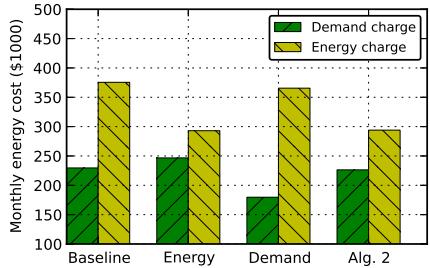}
	\vspace{-6mm}
	\caption{Cost breakdown comparison for geo-distributed data centers.}
	\label{fig:geo-breakdown}
	\end{minipage}
	\begin{minipage}[t]{0.33\linewidth}
	\centering
	\includegraphics[width=1\linewidth]{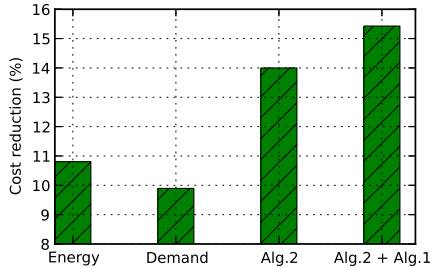}
	\vspace{-6mm}
	\caption{Cost reduction compared to {\sf Baseline}.}
	\label{fig:geo-reduction}
	\end{minipage}
	\begin{minipage}[t]{0.33\linewidth}
	\centering
	\includegraphics[width=1\linewidth]{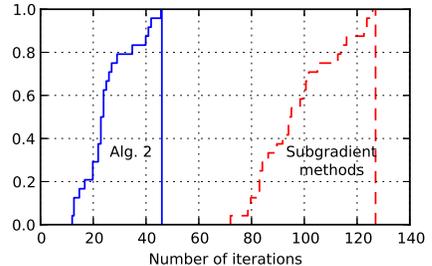}
	\vspace{-6mm}
	\caption{CDF of number of iterations.}
	\label{fig:geo-cdf}
	\end{minipage}
	\vspace{-3mm}
\end{figure*}

We now look at the case of multiple geo-distributed data centers, and examine more closely the cost savings from optimizing request routing, and the performance of Algorithm~\ref{alg:ours}. To do this we turn off partial execution in this set of simulation. We have three benchmarks here. The first, called {\sf Baseline}, directs user requests to the closest data center as long as capacity allows, and does not attempt to reduce energy cost. The second, called {\sf Energy}, optimizes request routing only for energy charge, i.e. it only considers the per kWh price and directs user requests to locations where the per kWh price is cheap while conforming to the average latency requirements. This represents a large class of existing works that shift workloads according to geographical diversity of the energy prices \cite{QWBG09,RLXL10,LLWL11,LWAT11,GCWK12,icac13,sigmetrics13}. On the other hand, the third, called {\sf Demand}, optimizes request routing only for demand charge, and tries to smooth out the demand patterns at locations where the per kW price is high. Finally, {\sf Alg. 2} refers to our Algorithm~\ref{alg:ours} that optimizes for both demand and energy charge, subject to the latency constraint.

Fig.~\ref{fig:geo-breakdown} shows the breakdowns of the total cost for all six data centers. Observe that the total cost stands around \$600K, with \$230K demand charge and \$380K energy charge as {\sf Baseline} shows. {\sf Energy} improves the situation by lowering the energy charge. However, it actually incurs a higher demand charge than {\sf Baseline}, as it shifts demands to locations with cheaper per kWh price where the per kW prices are not necessarily cheaper. Also the demand series are more fluctuating at those locations. Both factors contribute to the higher demand charge. {\sf Demand}, on the other hand, effectively reduces the demand charge, with only marginally reduced energy cost. By taking into account both factors, {\sf Alg. 2} offers the most cost savings as expected. In all cases, the latency constraint \eqref{cons:latency} is always satisfied. This confirms the benefits of request routing optimization for geo-distributed data centers. 

Fig.~\ref{fig:geo-reduction} further plots the percentage of cost savings provided by different schemes compared to {\sf Baseline}. {\sf Energy} and {\sf Demand} provide 10.8\% and 9.8\% cost savings, while {\sf Alg. 2} is able to offer 14\% cost savings. We also calculate the cost savings of joint request routing and partial execution by using Algorithm~\ref{alg:ours} together with Algorithm~\ref{alg:onedc}, shown as {\sf Alg.2 + Alg.1} in the figure. It provides 15.5\% cost reduction, amounting to around a monthly saving of \$85K for six data centers. Our results establish that our workload scheduling and request routing algorithms are effective in reducing the total energy cost for practical-scale data centers.

\subsection{Convergence}
\label{sec:convergence}

We now investigate the convergence and running time of our ADMM based Algorithm~\ref{alg:ours}. For comparison, we use the subgradient method \cite{BM} to solve the dual problem of the transformed optimization \eqref{opt:new_routing} with the augmented Lagrangian \eqref{eqn:our_lag}. Specifically, the primal variables $\alpha$ and $\beta$ are jointly optimized instead of sequentially updated as in our ADMM algorithm, and the dual variables $\lambda$ are updated by the subgradient method. The step size is carefully chosen according to the diminishing step size rule \cite{BM}. 

Fig.~\ref{fig:geo-cdf} plots the CDF of the number of iterations the two algorithms take to achieve convergence for the 30 runs on the traces. Our ADMM algorithm converges much faster than the subgradient methods. Our algorithm takes at most 46 iterations to converge in the worse case, and for 80\% of the time converges within 33 iterations. The subgradient method takes at least 72 iterations to converge, and for 80\% of the time takes more than 110 iterations. This shows the fast convergence of our ADMM algorithm compared to conventional methods.


%% file: related.tex
\section{Related Work}
\label{sec:related}

Many related works on thermal management and workload shifting to reduce data center energy cost have been discussed in Sec.~\ref{sec:intro}. Some other efforts include dynamically shutting down and waking up idle servers \cite{LWAT11}, using battery and/or on-site generators to absorb workload spikes \cite{GWSU13,TLCS13}, etc. These proposals are orthogonal to our approach using partial execution. A recent work \cite{LWCR13} focuses on the coincidental peak charge which is a form of demand response programs voluntary for data centers to participate to help better balance the grid. 

For partial execution, besides those discussed in Sec.~\ref{sec:partial_bk}, \cite{BC10} develops a flexible system that allows many programs to take advantage of approximation opportunities in a systematic manner to reduce energy. This enables the general implementation of partial execution while we focus more on the algorithmic challenges brought by partial execution and demand charge.

%% file: conclusion.tex
\section{Conclusion}
\label{sec:conclusion}

We proposed to use partial execution to reduce the peak power demand and total energy cost of data centers, given the importance of demand charge as established by our empirical study of real-world electricity contracts. We studied the resulting workload scheduling problem with SLA constraints in detail. The case with a single data center can be optimally solved. For geo-distributed data centers, we tackled the large-scale joint optimization of request routing and workload scheduling following a decoupling approach. Request routing is solved using an efficient distributed algorithm based on ADMM that decomposes the global problem into many sub-problems, each of which can be quickly solved. Trace-driven simulations are conducted to evaluate the algorithm's performance. As future work, we plan to more thoroughly study the impact of partial execution on demand response mechanisms of data centers.